\documentclass[aps,prl,twocolumn,showpacs,preprintnumbers,
               nofootinbib,float,superscriptaddress]{revtex4-1}

\usepackage{epsfig}
\usepackage{color}
\usepackage[dvipsnames]{xcolor}
\usepackage{float,amsmath,amssymb}
\usepackage{graphicx}
\usepackage{url}
\usepackage{bbold}
\usepackage[utf8]{inputenc}
\def\deltan{\hat\delta}
\usepackage{hyperref}

\begin{document}

\title{Observable signatures of initial state momentum anisotropies in nuclear collisions}


\author{Giuliano Giacalone}
\affiliation{Universit\'e Paris Saclay, CNRS, CEA, Institut de physique th\'eorique, 91191 Gif-sur-Yvette, France}

\author{Bj\"orn Schenke}
\affiliation{Physics Department, Brookhaven National Laboratory, Upton, NY 11973, USA}

\author{Chun Shen}
\affiliation{Department of Physics and Astronomy, Wayne State University, Detroit, Michigan 48201, USA}
\affiliation{RIKEN BNL Research Center, Brookhaven National Laboratory, Upton, NY 11973, USA}

\begin{abstract}
We show that the correlation between the elliptic momentum anisotropy, $v_2$, and the average transverse momentum, $[p_T]$, at fixed multiplicity in small system nuclear collisions carries information on the origin of the observed momentum anisotropy. A calculation using a hybrid IP-Glasma+\textsc{Music}+UrQMD model that includes contributions from final state response to the initial geometry as well as initial state momentum anisotropies of the Color Glass Condensate, predicts a characteristic sign change of the correlator $\hat{\rho}(v_2^2,[p_T])$ as a function of charged particle multiplicity in p+Au and d+Au collisions at $\sqrt{s}=200\,{\rm GeV}$, and p+Pb collisions at $\sqrt{s}=5.02\,{\rm TeV}$. This sign change is absent in calculations without initial state momentum anisotropies.
The model further predicts a qualitative difference between the centrality dependence of $\hat{\rho}(v_2^2,[p_T])$ in Au+Au collisions at $\sqrt{s}=200\,{\rm GeV}$ and Pb+Pb collisions at $\sqrt{s}=5.02\,{\rm TeV}$, with only the latter showing a sign change in peripheral events. Predictions for O+O collisions at different collision energy show a similar behavior.
Experimental observation of these distinct qualitative features of $\hat{\rho}(v_2^2,[p_T])$ in small and large systems would constitute strong evidence for the presence and importance of initial state momentum anisotropies predicted by the Color Glass Condensate effective theory.
\end{abstract}

\maketitle


\section{Introduction}
\vspace{-0.25cm}
Anisotropies observed in the transverse momentum distribution of long range rapidity correlations of charged hadrons, produced in collisions of small systems at the Relativistic Heavy Ion Collider (RHIC) and the Large Hadron Collider (LHC), can originate from final state response to the initial transverse geometry \cite{Bozek:2011if,Bozek:2012gr,Bozek:2013df,Bozek:2013uha,Bozek:2013ska,Bzdak:2013zma,Qin:2013bha,Werner:2013ipa,Kozlov:2014fqa,Schenke:2014zha,Romatschke:2015gxa,Shen:2016zpp,Weller:2017tsr,Mantysaari:2017cni}, as well as initial momentum anisotropies in the produced partons, which has been explicitly demonstrated within the Color Glass Condensate (CGC) effective theory \cite{Dumitru:2008wn,Kovner:2010xk,Dumitru:2010iy,Kovner:2011pe,Dusling:2012iga,Levin:2011fb,Dusling:2012wy,Dusling:2013qoz,Dumitru:2014dra,Dumitru:2014yza,Schenke:2015aqa,McLerran:2015sva,Schenke:2016lrs,Dusling:2017dqg,Dusling:2017aot,Mace:2018vwq,Mace:2018yvl,Kovner:2018fxj}.

Experimental data on momentum anisotropies in small systems, including p+Au, d+Au, and $^3$He+Au collisions at RHIC \cite{Aidala:2017ajz,PHENIX:2018lia,Lacey:2020ime}, and p+Pb collisions at LHC \cite{Chatrchyan:2013nka,Aad:2013fja,Acharya:2018lmh,Aaboud:2018ves,Acharya:2019vdf} have been described fairly successfully with hydrodynamics based models \cite{Bozek:2012gr,Bozek:2013uha,Bozek:2013ska,Qin:2013bha,Werner:2013ipa,Kozlov:2014fqa,Romatschke:2015gxa,Shen:2016zpp,Weller:2017tsr}, while calculations including only initial state effects have so far failed to describe the experimental data's systematics with multiplicity and system size.

Calculations using the CGC \cite{McLerran:1994ni,McLerran:1994ka,Iancu:2003xm} based IP-Glasma \cite{Schenke:2012wb,Schenke:2012hg} initial conditions followed by viscous hydrodynamic evolution include components of both effects \cite{Bzdak:2013zma,Schenke:2014zha,Mantysaari:2017cni,Schenke:2019pmk,Schenke:2020mbo}. In Ref.\,\cite{Schenke:2019pmk} the correlation between the initial momentum anisotropy and the final elliptic flow was determined and shown to be significant (compared to the correlation of the elliptic flow with the initial transverse geometry) for $dN_{\rm ch}/d\eta\lesssim 10$ in the small systems at RHIC.

Until now, there has not been a clear way to experimentally distinguish the dominant origin of azimuthal anisotropies in small systems. In this letter, we demonstrate how the measurement of the correlation between the mean transverse momentum and the elliptic anisotropy as a function of multiplicity provides a means to identify the role of initial state momentum anisotropies. In particular, in p+A and d+A collisions, the correlator $\hat{\rho}(v_2^2,[p_T])$ \cite{Bozek:2016yoj}, where $v_2$ is the transverse momentum $p_T$-integrated elliptic anisotropy, and $[p_T]$ the mean transverse momentum in a given event, exhibits a sign change from positive to negative with increasing multiplicity, when an initial momentum anisotropy is present.
Excluding initial state momentum anisotropies in a purely geometry driven version of our model results in negative $\hat{\rho}(v_2^2,[p_T])$ for all multiplicities. We note that purely geometry driven models can also produce positive $\hat{\rho}(v_2^2,[p_T])$ in small systems \cite{Bozek:2020drh}, depending on the details of the initial geometry, but a sign change as described above has not been observed.

We further predict distinct differences between the centrality dependencies of $\hat{\rho}(v_2^2,[p_T])$ in O+O collisions at RHIC and LHC energies, as initial state effects play a more important role (in peripheral events) for lower collision energies. For the same reason, we predict that while $\hat{\rho}(v_2^2,[p_T])$ in Pb+Pb collisions at $\sqrt{s}=5.02\,{\rm TeV}$ turns negative with decreasing multiplicity, consistent with experimental data from the ATLAS collaboration \cite{Aad:2019fgl}, the same observable in $\sqrt{s}=200\,{\rm GeV}$ Au+Au collisions remains positive. Observing this behavior at RHIC, along with the predicted sign changes of $\hat{\rho}(v_2^2,[p_T])$ in the small systems, would constitute strong evidence for the presence and importance of initial state momentum anisotropies as predicted by the CGC effective theory.

\section{Observable and estimators}\vspace{-0.25cm}
We compute the correlator \cite{Schenke:2020uqq}
\begin{equation}\label{eq:v2PT}
    \hat{\rho}(v_2^2,[p_T])=\frac{\langle \deltan v_2^2 \,\deltan [p_T]\rangle}{\sqrt{\langle(\deltan v_2^2)^2\rangle\langle(\deltan [p_T])^2\rangle}}\,,
\end{equation}
where the event-by-event deviation of any observable $O$, $\delta O = O - \langle O \rangle$, at fixed multiplicity is defined as~\cite{Olszewski:2017vyg} 
\begin{align}
\deltan O &\equiv \delta O  - \frac{\langle \delta O \delta N \rangle }{ \sigma_N^2 } \delta N \,.
\end{align}
Here, $N$ is the multiplicity and $\sigma_N$ the variance of $N$ in a given centrality bin. The elliptic flow coefficient  $v_2=v_2\{2\}$ is obtained from two-particle correlations. We note that the observable in Eq.\,\eqref{eq:v2PT} is equivalent to the one defined in \cite{Bozek:2016yoj}, computed in narrow bins of multiplicity. 

We present calculations based on the hybrid model consisting of the IP-Glasma initial state, \textsc{Music} viscous relativistic hydrodynamics \cite{Schenke:2010nt,Schenke:2010rr,Schenke:2011bn}, and UrQMD hadronic transport \cite{Bass:1998ca,Bleicher:1999xi}. All details of the calculation and resulting bulk observables are described in \cite{Schenke:2020mbo}. We use the same set of events for the analyses presented here.

We compute two predictors for the $\hat{\rho}$-correlator. The first, $\hat{\rho}_{\rm est}(\varepsilon_2^2,[s])$, is based entirely on the initial geometry (determined at $\tau=0.1\,{\rm fm}$), using the initial spatial eccentricity 
\begin{equation}
    \mathcal{E}_2 = \varepsilon_2 e^{i 2 \psi_2} = \frac{\langle x^2-y^2\rangle + i\langle 2xy\rangle}{\langle x^2+y^2\rangle},
\end{equation}
with $\langle \cdot \rangle$ using an energy density $(e)$ weight, as estimator for $v_2$. The second, $\hat{\rho}_{\rm est}(\varepsilon_p^2,[s])$, uses the initial momentum anisotropy as estimator for $v_2$. The momentum anisotropy can be computed from the initial energy momentum tensor of the IP-Glasma model as \cite{Schenke:2019pmk}
\begin{equation}\label{eq:epsilonp}
    \mathcal{E}_p \equiv \varepsilon_p e^{i 2 \psi_2^p} \equiv \frac{\langle T^{xx}-T^{yy}\rangle + i\langle 2 T^{xy}\rangle}{\langle T^{xx}+T^{yy}\rangle}\,,
\end{equation}
evaluated at $\tau =\,0.1{\rm fm}$, where here $\langle\cdot\rangle$ is defined without any weight. In both $\mathcal{E}_2$ and $\mathcal{E}_p$, we estimate $[p_T]$ from the average initial entropy density, $[s]=[e^{3/4}]$, which has been shown to be an optimal predictor of the average transverse momentum in small systems~\cite{Schenke:2020uqq}.

To confirm that agreement of $\hat{\rho}(v_2^2,[p_T])$ with one or the other estimator is rooted in the dominance of initial state or final state anisotropies for determining the observable $v_2$, we will also compute the Pearson coefficients of $\mathcal{E}_2$ with $V_2$ and $\mathcal{E}_p$ with $V_2$, defined as \cite{Gardim:2011xv,Gardim:2014tya,Betz:2016ayq}
\begin{equation}
    Q(\mathcal{E}, V_2) = \frac{{\rm Re} \langle\mathcal{E} V_2^*\rangle}{\sqrt{\langle |\mathcal{E}|^2\rangle \langle |V_2|^2\rangle}}\,,
\end{equation}
where $V_2$ is the complex valued $2^{\rm nd}$ order flow harmonic, here computed without $p_T$ cut. 

\section{Results}
\begin{figure}[t]
  \centering
  \includegraphics[width=0.48\textwidth]{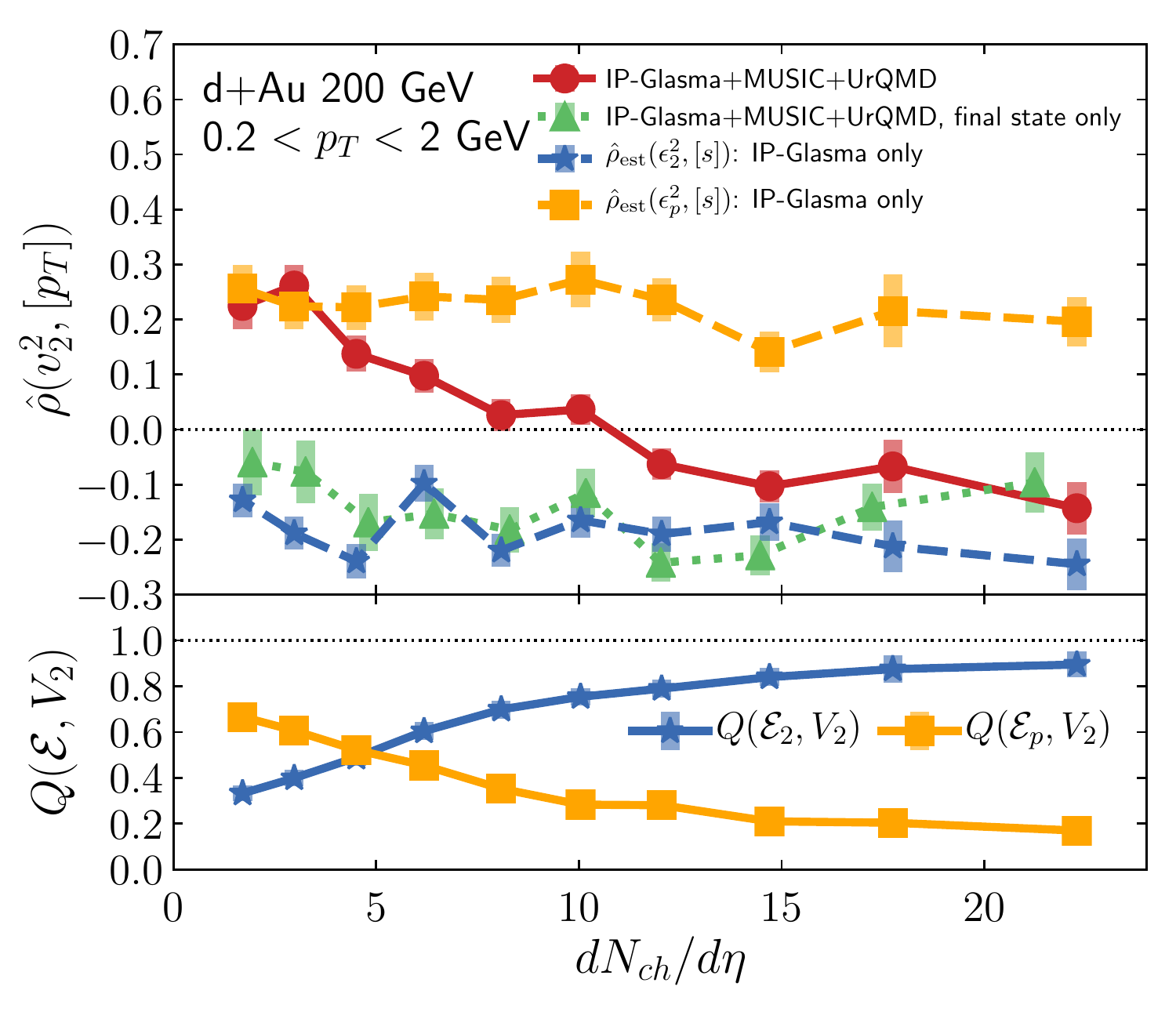}
  \caption{Upper panel: The correlator $\hat{\rho}(v_2^2,[p_t])$ (circles, solid lines) together with estimators based on the initial geometry ($\hat{\rho}_{\rm est}(\varepsilon_2^2,[s])$, stars) and the initial momentum anisotropy ($\hat{\rho}_{\rm est}(\varepsilon_p^2,[s])$, squares) in d+Au collisions at $\sqrt{s}=200\,{\rm GeV}$. Lower panel: Pearson coefficients between $v_2$ and the initial ellipticity (stars) and the initial momentum anisotropy (squares), respectively.  \label{fig:Pearson-rho-dAu}}
\end{figure}

\vspace{-0.25cm}
We present our main result in Fig.\,\ref{fig:Pearson-rho-dAu}, where in the upper panel we show $\hat{\rho}(v_2^2,[p_T])$ for $200\,{\rm GeV}$ d+Au collisions together with the geometric estimator $\hat{\rho}(\varepsilon_2^2,[s])$ and the estimator based on the initial momentum anisotropy $\hat{\rho}(\varepsilon_p^2,[s])$. We determine $\hat{\rho}(v_2^2,[p_T])$ in the transverse momentum range $0.2\,{\rm GeV}< p_T < 2\,{\rm GeV}$.
One can clearly see that for higher multiplicities $\hat{\rho}(v_2^2,[p_T])$ approaches the geometric estimator, while at lower multiplicities the initial momentum anisotropy predicts $\hat{\rho}(v_2^2,[p_T])$ better.

\begin{figure*}[t]
  \centering
  \includegraphics[width=0.48\textwidth]{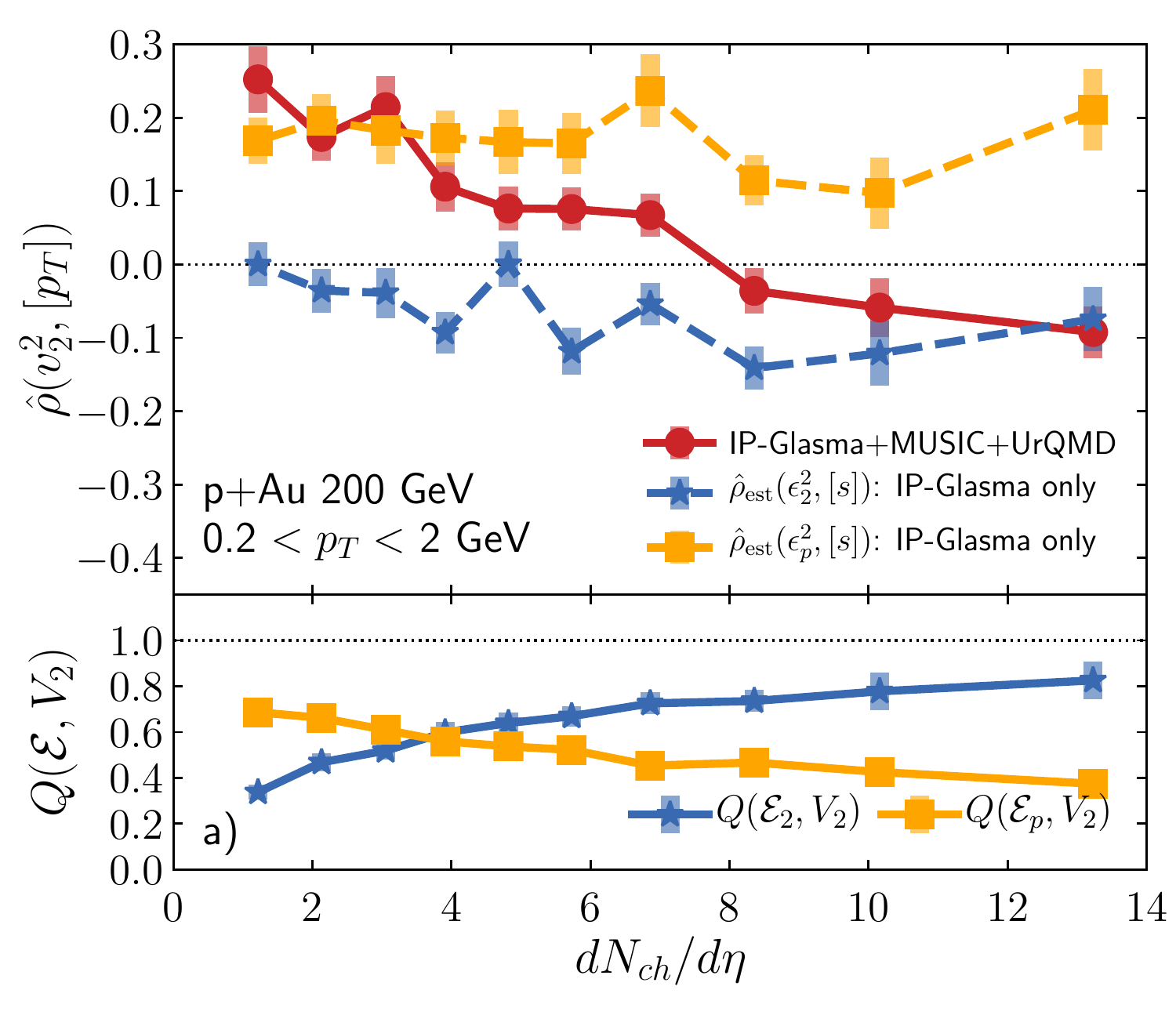}
  \includegraphics[width=0.48\textwidth]{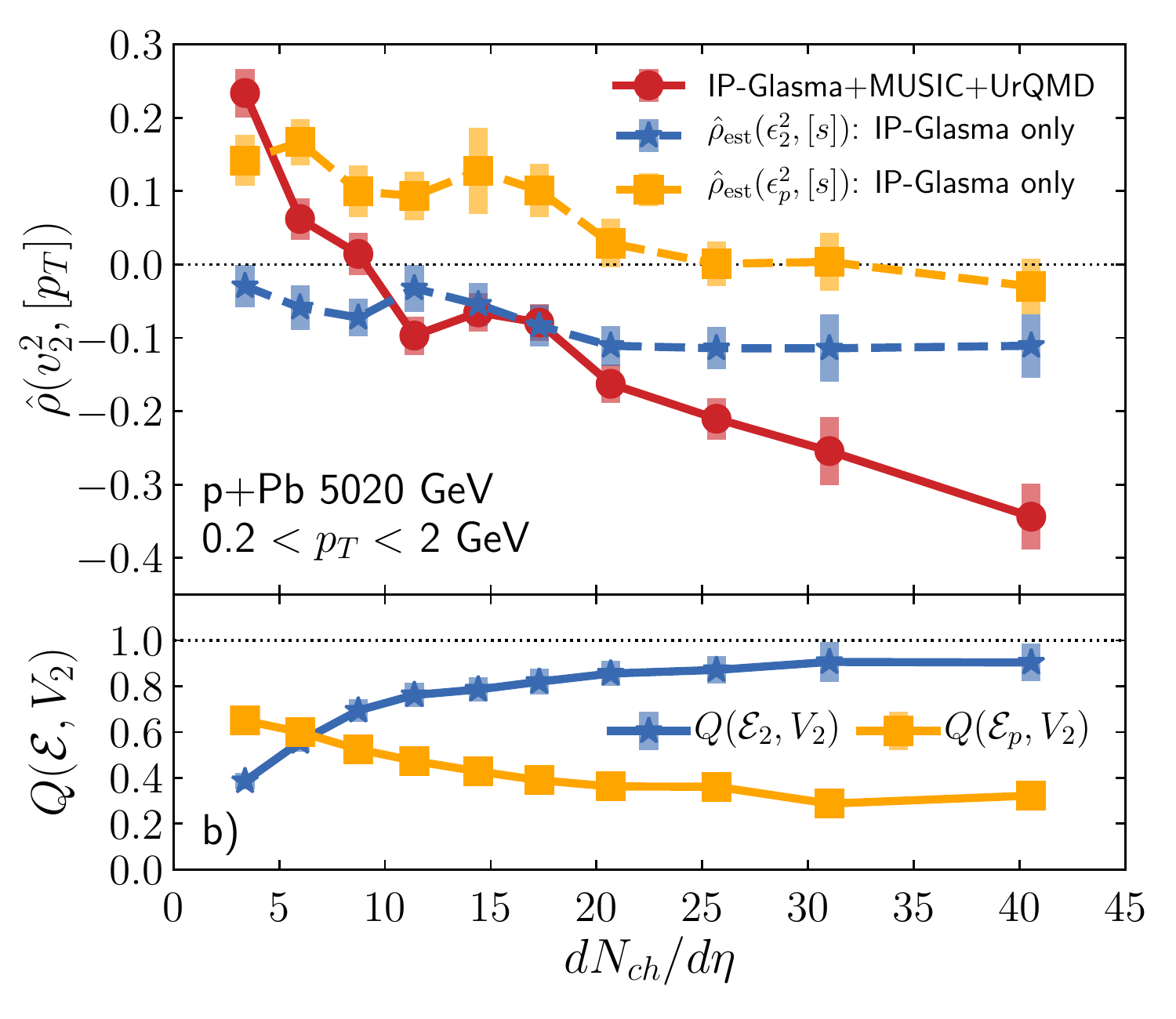}
  \includegraphics[width=0.48\textwidth]{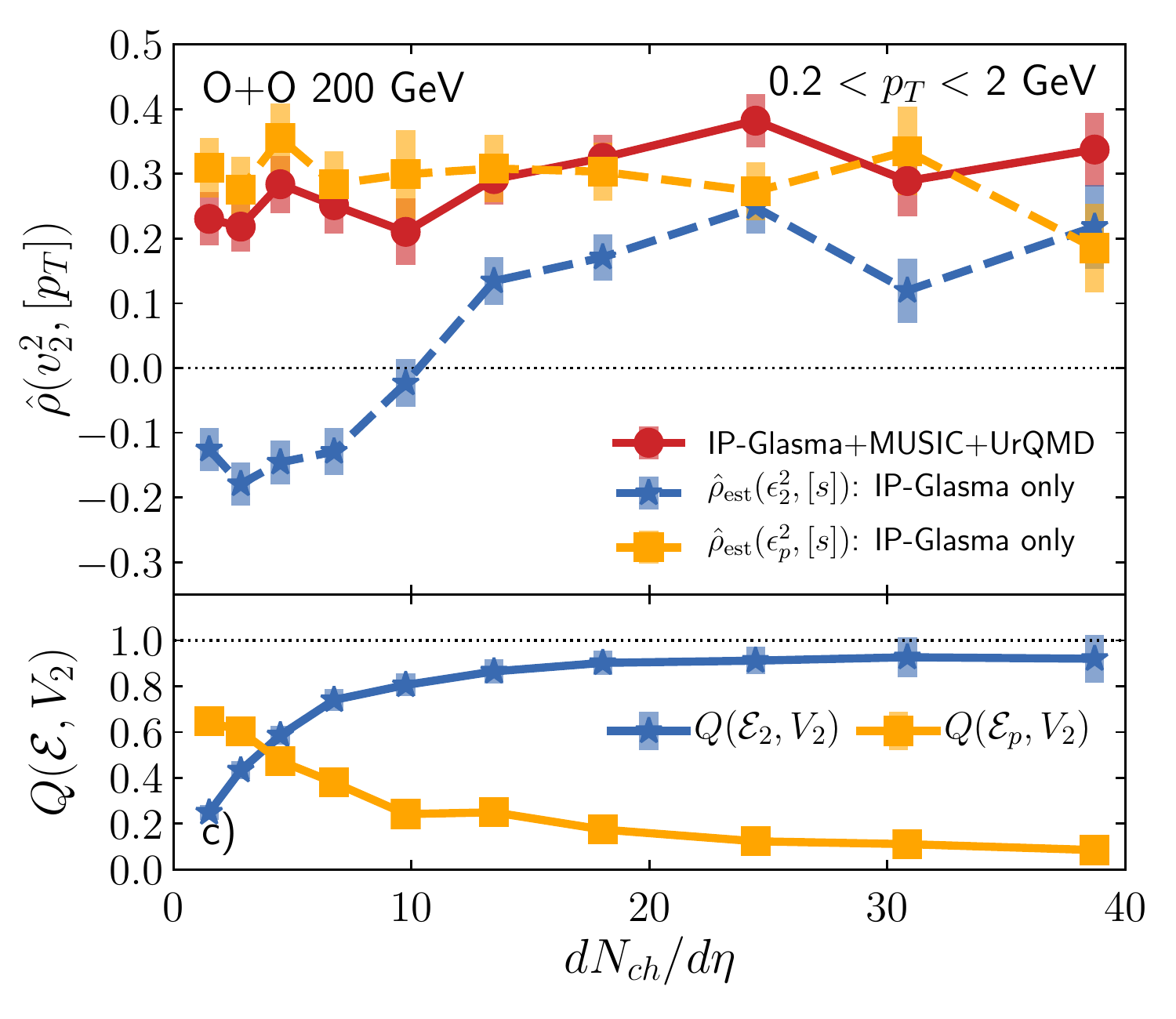}
  \includegraphics[width=0.48\textwidth]{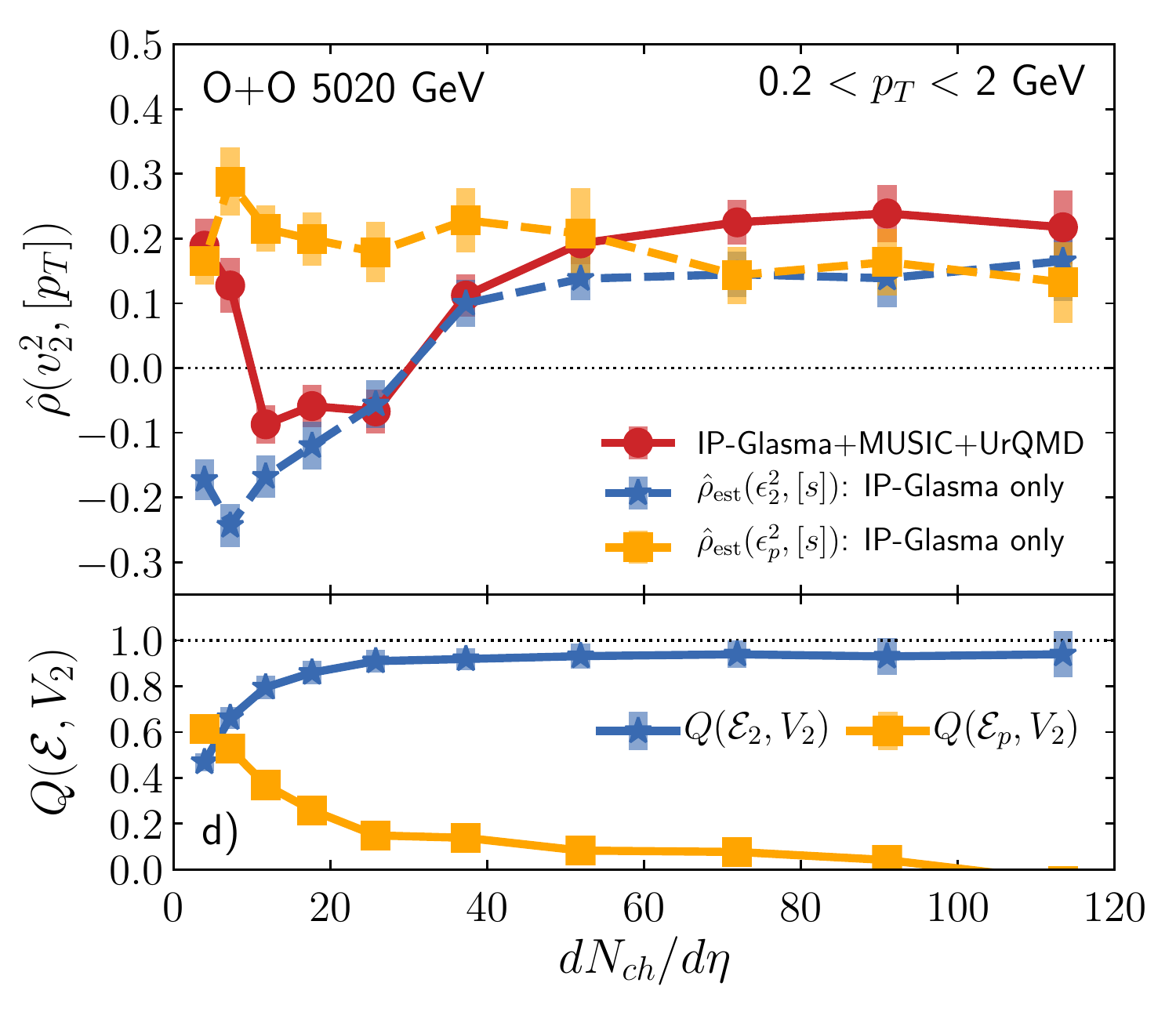}
  \caption{The correlator $\hat{\rho}(v_2^2,[p_t])$ (circles) together with estimators based on the initial geometry ($\hat{\rho}_{\rm est}(\varepsilon_2^2,[s])$, stars) and the initial momentum anisotropy ($\hat{\rho}_{\rm est}(\varepsilon_p^2,[s])$, squares) in a)  $\sqrt{s}=200\,{\rm GeV}$ p+Au, b) $\sqrt{s}=5.02\,{\rm TeV}$ p+Pb, c) $\sqrt{s}=200\,{\rm GeV}$, and d) $\sqrt{s}=5.02\,{\rm TeV}$ O+O collisions. Lower panels show the Pearson coefficients between $v_2$ and the initial ellipticity (stars) and the initial momentum anisotropy (squares), respectively. \label{fig:rho-pAu-pPb-OO}}
\end{figure*}

Based on the color domain interpretation of the initial state momentum anisotropy \cite{Kovner:2010xk,Kovner:2011pe,Dumitru:2014dra,Dumitru:2014vka,Lappi:2015vta}, we expect $\hat{\rho}_{\rm est}(\varepsilon_p^2,[s])$ to be positive, because at fixed multiplicity, a larger $[p_T]$ selects events with smaller transverse size. This reduces the number of color domains with an average size of $1/Q_s$, which enhances the magnitude of initial momentum anisotropy in the CGC description \cite{Lappi:2015vta}.

The Pearson coefficients $Q(\mathcal{E}, V_2)$ in the lower panel show that the behavior of $\hat{\rho}(v_2^2,[p_T])$ is a result of the geometry dominating the elliptic flow in high multiplicity events, and the initial momentum anisotropy driving the final $v_2$ at low multiplicity.\footnote{We note that $Q(\mathcal{E}_p,\mathcal{E}_2)$ ranges from consistent with zero within our statistical errors in central d+Au events to small negative values ($Q(\mathcal{E}_p,\mathcal{E}_2)\approx -0.06 \pm 0.02$) in more peripheral d+Au events.} 
The Pearson coefficients were studied already in \cite{Schenke:2019pmk}, but they are not experimentally observable. In contrast, with $\hat{\rho}(v_2^2,[p_T])$ we have now identified an observable whose sign change as a function of multiplicity is a clean indicator of the origin of the elliptic flow in small systems and the presence of initial state momentum anisotropies as predicted from the Color Glass Condensate. 

To further support this statement, we also show results obtained from a calculation that only uses the initial energy density of the IP-Glasma calculation, and starts the hydrodynamic evolution at $\tau=0.1\,{\rm fm}$ to compensate for the lack of initial radial flow. In this case, there is no initial momentum anisotropy and the only source of elliptic flow is geometry driven. The resulting $\hat{\rho}(v_2^2,[p_T])$, shown as triangles in Fig.\,\ref{fig:Pearson-rho-dAu}, is qualitatively different, as no sign change is present. In fact, $\hat{\rho}(v_2^2,[p_T])$ follows the geometric predictor $\hat{\rho}(\varepsilon_2^2,[s])$ well over the entire range of multiplicity. 

We conclude that the experimental observation of a sign change of $\hat{\rho}(v_2^2,[p_T])$ from positive to negative with increasing multiplicity in d+Au collisions at $\sqrt{s}=200\,{\rm GeV}$ will be evidence for CGC initial state momentum anisotropies, that have so far eluded observation.

We show results for different systems and energies in Fig.\,\ref{fig:rho-pAu-pPb-OO}. In panel a) we present our prediction for $200\,{\rm GeV}$ p+Au collisions, where again a sign change of $\hat{\rho}(v_2^2,[p_T])$ is clearly visible. The same holds for $5.02\,{\rm TeV}$ p+Pb collisions, shown in panel b), with the sign change occurring at a similar value of $dN_{\rm ch}/d\eta$. Unfortunately, existing p+Pb collision data from the ATLAS Collaboration \cite{Aad:2019fgl} does not go down to low enough $dN_{\rm ch}/d\eta$ to confirm the presence of this feature.

\begin{figure*}[t]
  \centering
  \includegraphics[width=0.49\textwidth]{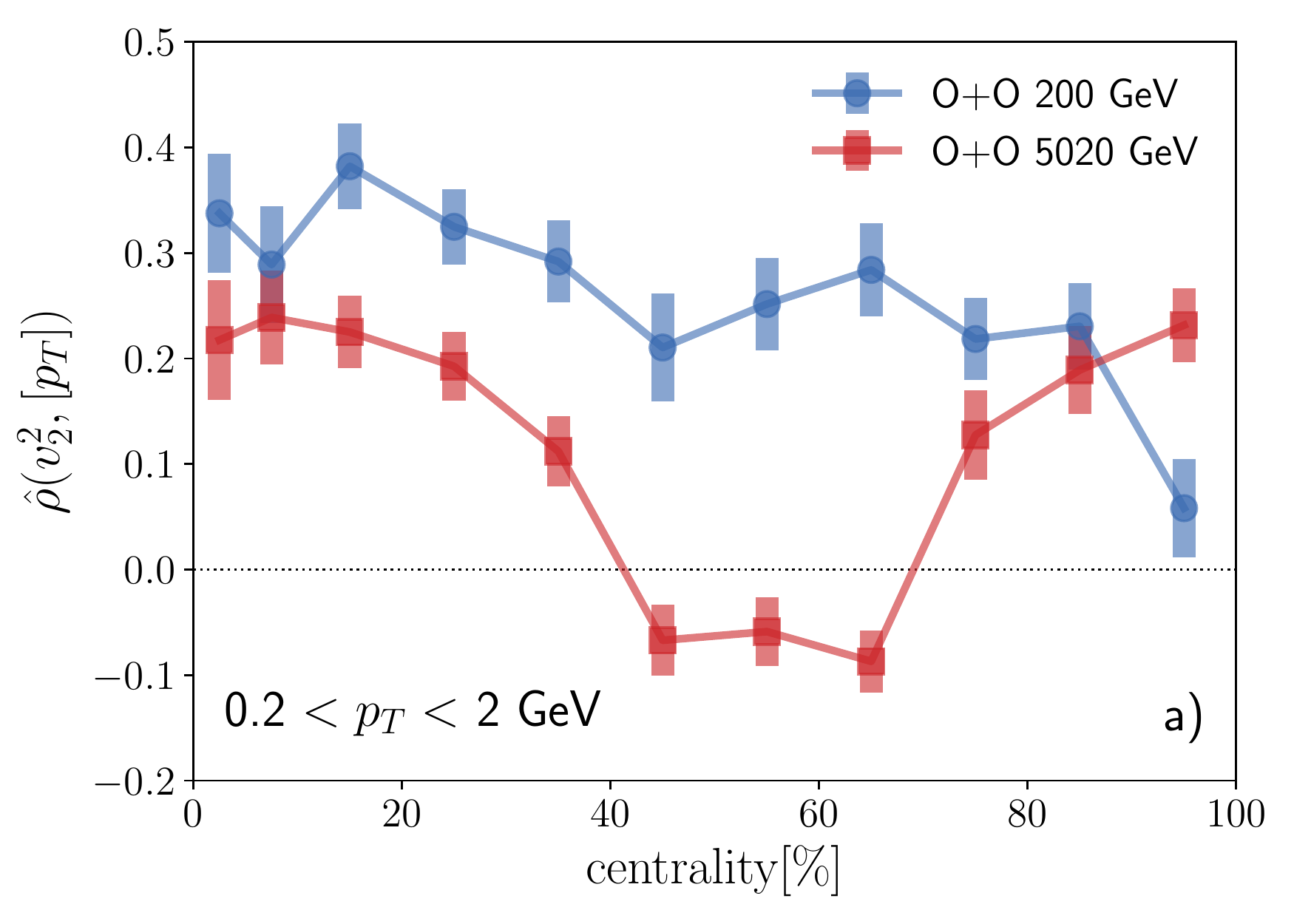}
  \includegraphics[width=0.49\textwidth]{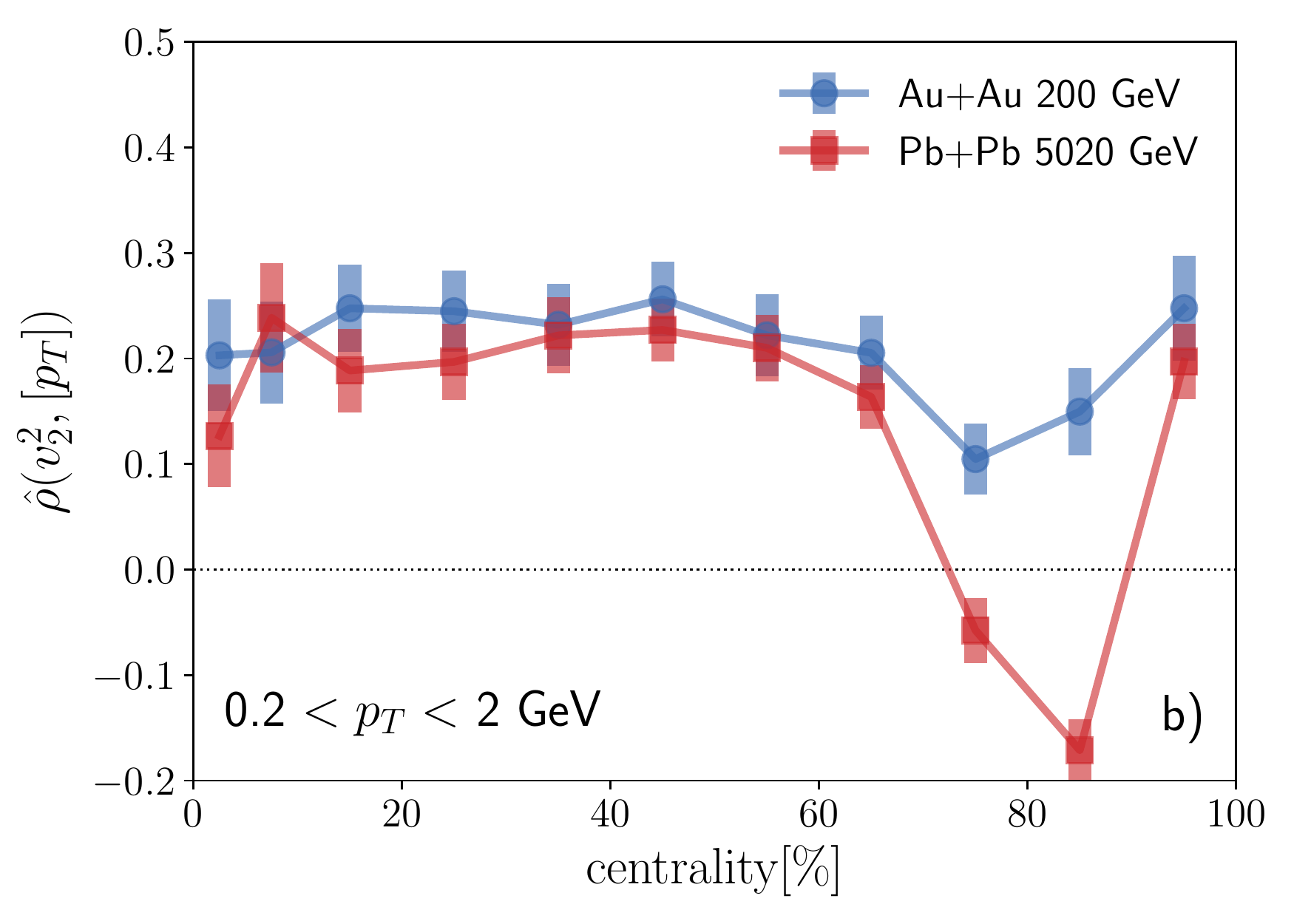}
  \caption{The correlator $\hat{\rho}(v_2^2,[p_t])$ in a) O+O collisions for $\sqrt{s}=200\,{\rm GeV}$ (circles) and $\sqrt{s}=5.02\,{\rm TeV}$ (squares), and in b) $\sqrt{s}=200\,{\rm GeV}$ Au+Au (circles) and $\sqrt{s}=5.02\,{\rm TeV}$ Pb+Pb (squares) collisions as functions of centrality. Results at the higher collision energy exhibit two sign changes. \label{fig:rho-cent}}
\end{figure*}

The energy dependence is best studied when using the same type of collision system at both energies. This would be achievable with future O+O runs at both RHIC and LHC. Comparing $200\,{\rm GeV}$ (c) and $5.02\,{\rm TeV}$ O+O collisions (d), a clear qualitative difference is visible. While at the lower collision energy $\hat{\rho}(v_2^2,[p_T])$ is always positive, it changes sign twice at the higher collision energy. This behavior can be interpreted as final state effects being more dominant at the higher collision energy, where the lifetime of the system is longer. 

Fig.\,\ref{fig:rho-cent} a) shows the same results for $\hat{\rho}(v_2^2,[p_T])$ in O+O collisions at the two different collision energies as a function of centrality, emphasizing the qualitative difference between the two. Fig.\,\ref{fig:rho-cent} b) shows the same observable as a function of centrality for $200\,{\rm GeV}$ Au+Au collisions and $5.02\,{\rm TeV}$ Pb+Pb collisions, two systems for which data has already been taken. We predict that the sign change of $\hat{\rho}(v_2^2,[p_T])$ observed in peripheral Pb+Pb collisions by the ATLAS Collaboration~\cite{Aad:2019fgl}, and also reported at both RHIC and LHC energy by theoretical calculations that include only final state effects~\cite{Giacalone:2020dln,Giacalone:2020awm}, will not be observed in peripheral Au+Au collisions at RHIC energy. Experimental confirmation of this result will be a strong indication of an important contribution from the initial state momentum anisotropy to the observed $v_2$ in peri\-pheral events.

\vspace{-0.25cm}
\section{Conclusions}\vspace{-0.25cm}
We have demonstrated a way to experimentally observe the initial momentum anisotropy from the Color Glass Condensate using measurements of the correlation between the elliptic momentum anisotropy and the mean transverse momentum in small systems.

The correlator $\hat{\rho}(v_2^2,[p_T])$, computed in a framework including both final state effects and initial state momentum anisotropies shows a sign change as a function of multiplicity in $200\,{\rm GeV}$ p+Au and d+Au collisions at RHIC and $5.02\,{\rm TeV}$ p+Pb collisions at LHC. The sign change is not present in simulations that include only geometry-driven final state effects, which demonstrates the robustness of our conclusion. We leave a more systematic study, in particular regarding the role of the kinematic cuts, to a follow-up work.

We further predict that in the presence of initial state momentum anisotropies, the correlator $\hat{\rho}(v_2^2,[p_T])$ is positive for all multiplicities in $200\,{\rm GeV}$ O+O and Au+Au collisions, while it changes sign in $5.02\,{\rm TeV}$ O+O and Pb+Pb collisions. Experimental observation of these clean qualitative signatures in peripheral heavy ion and small system collisions will be the first evidence for the presence and importance of initial state momentum anisotropies predicted by an effective theory of QCD. This will establish that the study of high energy nuclear collisions in the regime  $dN_{\rm ch}/d\eta \lesssim 40$ provides the rare opportunity to study the detailed properties of a non-Abelian theory in the laboratory.

\vspace{-0.25cm}
\section*{Acknowledgments}\vspace{-0.25cm}
We thank Matt Luzum, Derek Teaney and Prithwish Tribedy for helpful discussions.
B.P.S. is supported under DOE Contract No.~DE-SC0012704. C.S. is supported under DOE Contract No.~DE-SC0013460. This research used resources of the National Energy Research Scientific Computing Center, which is supported by the Office of Science of the U.S. Department of Energy under Contract No.~DE-AC02-05CH11231 and resources of the high performance computing services at Wayne State University. This work is supported in part by the U.S. Department of Energy, Office of Science, Office of Nuclear Physics, within the framework of the Beam Energy Scan Theory (BEST) Topical Collaboration.

\vspace{-0.25cm}
\bibliography{spires}
\end{document}